\documentclass[dvips]{arxstspdf}
\usepackage{flushend}
\usepackage{stfloats}

\volume{26}
\issue{2}
\pubyear{2011}
\firstpage{162}
\lastpage{174}
\doi{10.1214/10-STS318}

\makeatletter
\newproclaim{example}{Example}
\makeatother

\begin{document}
\begin{frontmatter}

\title{Calibrated Bayes, for Statistics in General, and Missing Data in Particular\thanksref{T1}}
\relateddoi{T1}{Discussed in
\doi{10.1214/10-STS318B} and
\doi{10.1214/10-STS318A};
rejoinder at \doi{10.1214/10-STS318REJ}.}
\runtitle{Bayesian Missing Data Methods}

\begin{aug}
\author[a]{\fnms{Roderick} \snm{Little}\corref{}\ead[label=e1]{rlittle@umich.edu}}
\runauthor{R. Little}

\affiliation{University of Michigan}

\address[a]{Roderick Little is Richard D. Remington Collegiate Professor,
Department of Biostatistics, University of Michigan, 1415 Washington Heights,
Ann Arbor, Michigan 48109, USA \printead{e1}.}

\end{aug}

\begin{abstract}
It is argued that the Calibrated Bayesian (CB) approach to statistical
inference capitalizes on the strength of Bayesian and frequentist
\mbox{approaches} to statistical inference. In the CB approach, inferences
under a particular model are Bayesian, but frequentist methods are
useful for model development and model checking. In this article the CB
approach is outlined. Bayesian methods for missing data are then
reviewed from a CB perspective. The basic theory of the Bayesian
approach, and the closely related technique of multiple imputation, is
described. Then applications of the Bayesian approach to normal models
are described, both for monotone and nonmonotone missing data patterns.
Sequential Regression Multivariate Imputation and Penalized Spline of
Propensity Models are presented as two useful approaches for relaxing
distributional assumptions.
\end{abstract}


\end{frontmatter}

\section{Introduction}

There was clearly a time, perhaps not too far in the recent past, when
Bayesian methods were considered ``beyond the pail'' by frequentist
statisticians. But Bayesian methods have been resurgent in recent years,
to the extent that few statisticians have no interest in them, even if
they do not buy the complete philosophical package.

In this article I summarize my perspective on the role of Bayesian
methods in statistics, borrowing from a more extensive discussion in
Little (\citeyear{Little2006}). I then provide a brief overview of Bayesian\vadjust{\eject} inference for
missing data problems, both modeling and ignoring the missing data
mechanism, and multiple imputation (MI), an important practical tool for
dealing with missing data that has a Bayesian etiology. Finally, I give
some examples of Bayesian missing-data methods which I believe
frequentists could profitably add to their analytical toolkit.

Bayesian methods are particularly useful for handling missing data
problems in statistics. Incomplete data problems are readily amenable to\vadjust{\goodbreak}
likelihood-based methods, since they do not require a rectangular data
matrix. Maximum likelihood (ML) is an important approach, but the
loglikelihoods corresponding to missing data problems are typically more
complex than likelihoods for complete data, deviate more from the
quadratic approximations that underlie asymptotic inferences, and are
subject to under-identification or weak identification of parameters.
Consequently, ML requires iterative calculations, information
matrix-based standard errors are often difficult to compute in
high-dimensional problems, and asymptotic ML inferences can have serious
deficiencies, particularly with small fragmentary samples. In contrast,
draws from the Bayesian posterior distribution can often be computed
using direct simulation or Markov Chain Monte Carlo techniques, and
these provide estimates of standard errors without the need to compute
and invert the information matrix. The inferences based on the posterior
distribution often have better frequentist properties than asymptotic
inferences based on ML. Furthermore, multiple imputations can be
generated as a byproduct of Bayesian calculations, and provide a~%
practically useful and flexible tool for solving missing data problems.
These points are further developed in the material that follows.

\section{Why Bayes? The Calibrated Bayes Philosophy}\label{sec:2}

The statistics world is still largely divided into frequentists, who
base inference for an unknown parameter $\theta$ on hypothesis tests or
confidence intervals \mbox{derived} from the distribution of statistics in
repeated sampling, and Bayesians, who formulate a model for the data and
prior distribution for unknown parameters, and base inferences for
unknowns on posterior distributions. Bayesians are also ``subjective,''
as when proper priors are elicited, and ``objective,'' as when
conventional ``reference priors'' are adopted. Both these facets of the
Bayesian paradigm have useful roles, depending on context. Asymptotic
maximum likelihood inference can be seen as a form of large sample
Bayes, with the interval for $\theta$ being interpreted as a posterior
credibility interval rather than a confidence interval.

I believe both systems of statistical inference have strengths and
weaknesses and, hence, the best course is to seek a compromise that
combines them in a way that capitalizes on their strengths. The Bayesian
paradigm is best suited for making statistical inference under an
assumed model. Indeed, under full probability modeling, with prior
distributions assigned to parameters, the Bayes theorem is indeed a~%
theorem---the path determined by probability theory to inference about
unknowns given the data, whether the targets are parameters or
predictions of unobserved quantities.

The frequentist approach, on the other hand, has various well-known
limitations regarding inference under an assumed model. First, it is not
prescriptive: frequentist theory seems more like a set of concepts for
assessing properties of inference procedures, ra\-ther than an inferential
system \textit{per se}. Under an agreed model, and assuming large
samples, there is a relatively prescriptive path to inference based\vadjust{\eject} on
maximum likelihood (ML) estimates and their large-sample distribution.
However, other frequentist approaches are entertained in practice, like
generalized estimating equations, based on robustness or other
considerations. Also, there is no prescriptive frequentist approach to
small sample problems. Indeed, for many problems, such as the
Behrens--Fisher problem of comparing two means of normal distributions
with different unknown variances, no procedure exists that has exact
repeated-sampling properties, such as exact nominal confidence coverage
for all values of the unknown parameter. Baye\-sian methods provide exact
frequentist coverage for some complete-data
problems---this occurs, in
particular, in problems where the Bayesian and frequentist inferences
are the same, as in $t$ inference for normal multiple regression with a
uniform prior on the regression coefficients and log variance. For more
complex problems, including problems with missing data, Bayesian methods
do not generally provide exact frequentist coverage, but they often
improve on ML by providing better small-sample inferences, perhaps
because Bayesian model shrinkage moderates inferences based on extreme
parameters estimates. As just one example, consider the adjustment of
estimates for categorical data motivated by Bayesian ideas (Agresti, \citeyear{Agresti2002}).

The Bayesian approach is prescriptive in the sense that, once a model
and prior distribution are specified, there is a clear path to
inferences based on the posterior distribution, or optimal estimates for
a given choice of loss function. There is no prescription for choosing
the model and prior distribution---that is what makes applied
statistics interesting---but certain ``reference'' prior distributions
for com\-plete-data problems can be expected to yield good frequentist
properties when applied to missing data problems; see, for example,
Little (\citeyear{Little1988}).

Frequentist inference violates the likelihood principle, and is
ambiguous about whether to condition on ancillary or approximately
ancillary statistics when performing repeated sampling calculations.
Little (\citeyear{Little2006}) provides more discussion, with
examples.

An attractive feature of Bayesian methods in com\-plete-data or
missing-data problems is the treatment of nuisance parameters. Bayesian
inference integrates over these parameters, rather than fixing them at
their ML estimates. This tends to yield inferences with improved
frequentist properties, since the uncertainty about these parameters is
taken into\vadjust{\eject} account. For example, for complete or incomplete data
problems, restricted ML, which integrates over location parameters, is
generally viewed as superior to ML, which maximizes them.

If we were handed the model on a plate and told to do inference for
unknowns, then Bayesian statistics is the clear winner. The problem, of
course, is that we never know the true model. Bayesian inference
requires and relies on a high degree of model speci\-fication
(Efron, \citeyear{Efron1986})---full specification of a~likeli\-hood and prior. Developing a good
model is challenging, particularly in complex problems. Furthermore, all
models are wrong, and bad models lead to bad answers: under the
frequentist paradigm, the search for procedures with good frequentist
proper\-ties provides some degree
of protection against model
misspecification, but there seems no such built-in protection under a
strict Bayesian paradigm where frequentist calculations are not
entertained.

Good principles for picking models are essential, and here I feel
frequentist methods have an important role. We want models that yield
inferences with good frequentist properties, such as 95\% credibility
intervals that cover the unknown parameter 95\% of the time if the
procedure was applied to repeated samples. The Bayesian has some tools
for model development and checking, like Bayes factors and model
averaging, but Bayesian hypothesis testing has well known problems, and,
in my view, frequentist approaches are essential when it comes to model
development and assessment.

To summarize, Bayesian statistics is strong for inference under an
assumed model, but relatively weak for the development and assessment of
models. Frequentist statistics provides useful tools for model
development and assessment, but has weaknesses for inference under an
assumed model. If this summary is accepted, then the natural compromise
is to use frequentist methods for model development and assessment, and
Bayesian methods for inference under a model. This capitalizes on the
strengths of both paradigms, and is the essence of the approach known as
Calibrated Bayes (CB).

Many statisticians have advanced CB ideas (e.g., Peers, \citeyear{Peers1965};
Welch, \citeyear{Welch1965};
Dawid, \citeyear{Dawid1982}), but I was particularly influenced by seminal papers
by Box (\citeyear{Box1980}) and Rubin (\citeyear{Rubin1984}). Box (\citeyear{Box1980}) wrote,

\begin{quote}
``I believe that$\ldots$ sampling theory is needed for exploration and
ultimate criticism of the entertained model in the light of the current
data, while Bayes' theory is needed for estimation of parameters
conditional on adequacy of the model.''
\end{quote}

\noindent He based his implementation of this idea on the factorization:
\[
p(Y,\theta |\mathrm{Model}) = p(Y|\mathrm{Model})p(\theta
|Y,\mathrm{Model}),
\]
where the second term on the right side is the posterior distribution of
the parameter $\theta$ given data~$Y$ and Model, and is the basis for
inference, and the first term on the right side is the marginal
distribution of the data $Y$ under the Model, and is used to assess the
validity of the Model, with the aid of frequentist considerations.
Specifically, discrepancy functions of the observed data $d(Y)$ are
assessed from the perspective of realizations from their marginal
distribution $p(d(Y)|\mathrm{Model})$. A questionable feature of this
``prior predictive checking'' is that checks are sensitive to the choice
of prior distribution even when this choice has limited impact on the
posterior inference; in particular, it leads to problems with assessment
of models involving noninformative priors.

Rubin (\citeyear{Rubin1984}) wrote,

\begin{quote}
``The applied statistician should be Baye\-sian in principle and
calibrated to the real world in practice---appropriate frequency
calculations help to define such a tie$\ldots$ frequency calculations are
useful for making Bayesian statements scientific, scientific in the
sense of capable of being shown wrong by empirical test; here the
technique is the calibration of Bayesian probabilities to the
frequencies of actual events.''
\end{quote}

\noindent Rubin (\citeyear{Rubin1984}) advocated model checking based on a different distribution,
namely, $p(Y^*,\theta^*|Y,\mathrm{Model})$, the predictive distribution
of future data $Y^*$ and parameter values $\theta^*$ given the Model and
observed data $Y$. This leads to posterior predictive checks (Rubin, \citeyear{Rubin1984};
Gelman, Meng and Stern, \citeyear{GelmanMengStern1996}), which extend frequentist checking
methods by not limiting attention to checking statistics that have a
known distribution under the model. These checks involve an amalgam of
Bayesian and frequentist ideas, but are clearly frequentist in spirit in
that they concern embedding the observed data within a sequence of
unobserved data sets that could have been generated under the Model, and
seeing whether the observed data are ``reasonable.''

Philosophy aside, perhaps the main reason why Bayesian methods have
flourished in recent years is\vadjust{\eject} the development of powerful computational
tools,~li\-ke the Gibbs' sampler and other Markov Chain Mon\-te Carlo (MCMC)
methods. These, together with gains in computing power, have made it
computationally feasible to carry out the high-dimensional integrations
required. An important early breakt\-hrough in MCMC methods actually
occurred for a missing data problem, as I discuss in Example~\ref{ex2} below.
Even if frequentists are completely against Bayes, they can apply these
Bayesian computational tools with weak prior distributions, and
interpret results as approximations to ML, with similar asymptotic
properties.

\section{A Short History of Statistical Analysis With Missing Data}\label{sec:3}

I divide the development of missing data methods in statistics into four
eras:

\subsection{Pre-EM Algorithm (Pre-1970s)}\label{sec31}

Early missing data methods
involved complete-case analysis, that is, simply discarding data with
any values missing, or simple imputation methods, which filled in
missing values with best estimates and analyzed the filled-in data. The
latter approach was developed quite extensively in the case of analysis
of variance with missing outcomes, which were imputed to maintain a
balanced design and hence an easily inverted design matrix (see Little
and Rubin, \citeyear{LittleRubin2002}, Chapter~2). These ingenious methods are now mainly of
historical interest, since inverting the design matrix corresponding to
the unbalanced data is not a big problem given advances in modern
computing. ML methods were developed for some simple missing data
problems, notably bivariate normal data with missing values on one
variable, which Anderson (\citeyear{Anderson1957}) solved noniteratively by factoring the
likelihood (see Example~\ref{ex1} below). ML for complex problems was iterative
and generally too hard given limits of computation, although progress
was made for contingency tables (Hartley, \citeyear{Hartley1958}) and normal models
(Hartley and Hocking, \citeyear{HartleyHocking1971}).

\subsection{The Maximum Likelihood Era\break (1970s--Mid 1980s)}\label{sec32}

ML
methods became popular and feasible in the mid-1970s with the
development of the Expectation--Maximization (EM) algorithm. EM builds a
link with complete-data ML and is simple to program in several important
multivariate models, including the multivariate normal model with a
general\vadjust{\eject} pattern of missing values. The term EM was coined in the famous
paper by Dempster, Laird and Rubin (\citeyear{DempsterLairdRubin1977}), which established some key
properties of the method, including the fact that the likelihood does
not decrease at each iteration. The EM algorithm had been previously
discovered several times for particular models (e.g., \mbox{McKendrick}, \citeyear{McKendrick1926};
Hartley, \citeyear{Hartley1958};
Baum et al., \citeyear{BaumEtAl1970}), and had been formulated in some
generality by Orchard and Woodbury (\citeyear{OrchardWoodbury1972}) and Sundberg (\citeyear{Sundberg1974}). The
simplicity and versatility of EM motivated extensions of EM to handle
more difficult problems, and applications to a~variety of complete-data
models for categorical and continuous data, as reviewed in Little and
Rubin (\citeyear{LittleRubin1987}), McLachlan and Krishnan (\citeyear{McLachlanKrishnan1997}) and Meng and van Dyk
(\citeyear{MengvanDyk1997}). For generalizations of the EM idea, see Lange (\citeyear{Lange2004}).

Another important development in this era was the formulation of models
for the missing data mechanism, and associated sufficient conditions for
when the missing data mechanism can be ignored for frequentist and
Bayesian inference (Rubin, \citeyear{Rubin1976}).

\subsection{Bayes and Multiple Imputation\break (Mid 1980s--Present)}\label{sec33}

The
transition from ML to Bayesian methods in the missing data setting was
initiated by Tanner and Wong (\citeyear{TannerWong1987}), who described data augmentation to
generate the posterior distribution of the parameters of the
multivariate normal model with missing data. Data augmentation is
closely related to the Gibbs' sampler, as discussed below. Another
important development was Rubin's (\citeyear{Rubin1978,Rubin1987,Rubin1996}) proposal to handle
missing data in public use data sets by multiple imputation (MI),
motivated by Bayesian ideas. In its infancy, this proposal seemed very
exotic and computationally impractical---not any more! MCMC facilitates
Bayesian multiple imputation, and is now implemented in
publicly-availa\-ble software for the convenience of users of both
Bayesian and frequentist persuasions.

\subsection{Robustness Concerns\break (1990s--Present)}\label{sec34}

Model-based
missing data methods are potentially vulnerable to model
misspecification, although they tend to outperform na\"{\i}ve methods
even when the model is misspecified (e.g., Little, \citeyear{Little1979}). Modern
interest in limiting effects of model misspecification by adopting
robust procedures has extended to missing data problems, notably with
``doubly robust'' procedures based on augmented inverse-probability
weighted estimating equations (Robins, Rotnitsky and Zhao, \citeyear{RobinsRotnitzkyZhao1994}). From a
more directly model-based Bayesian perspective, robustness takes the
form of developing models that make relatively weak structural
assumptions. A~method based on one such model, Penalized Spline of
Propensity Prediction (PSPP, Little and An, \citeyear{LittleAn2004}), is discussed in
Example~\ref{ex4} below. Another aspect of robustness concerns has been more
interest in model checks for standard missing data models (Gelman et
al., \citeyear{GelmanEtAl2005}).

I now sketch the likelihood and Bayesian theory for the analysis of data
with missing values that underlies the methods in Sections~\ref{sec32} and~\ref{sec33}.
I then describe the transition from Sections~\ref{sec32} to~\ref{sec34} for the case of
multivariate normal models, and elaborations for non-normal data.

\section{Likelihood-Based Methods With\break Missing Data}\label{sec:4}

Likelihoods can be defined for nonrectangular da\-ta, so likelihood
methods apply directly to missing-data problems:
\[
\mbox{statistical model} + \mbox{incomplete data} \Rightarrow \mbox{likelihood}.
\]
Given the likelihood function, standard approaches are to maximize it,
leading to ML, with associated large sample standard errors based on the
information, the sandwich estimator or the bootstrap; or to add a prior
distribution and compute the posterior distribution of the parameters.
Draws from the predictive distribution of the missing values can also be
created as a basis for MI.

As described in Little and Rubin (\citeyear{LittleRubin2002}, Chapter~6), let $Y = (y_{ij})_{n
\times K}$ represent a data matrix with~$n$ rows (cases) and $K$ columns
(variables), and define the missing-data indicator matrix $M =\break
(m_{ij})_{n \times K}$, with entries $m_{ij} = 0$ if $y_{ij}$ is
observed, $m_{ij} = 1$ if $y_{ij}$ is missing. Also, write $Y =
(Y_{\mathrm{obs}},Y_{\mathrm{mis}})$, where $Y_{\mathrm{obs}}$
represents the observed components of $Y$ and $Y_{\mathrm{mis}}$ the
missing components. A full parametric model factors the distribution of
($Y, M$) into a distribution $f(Y|\theta)$ for $Y$ indexed by unknown
parameters $\theta$, and a distribution $f(M|Y,\psi)$ for $M$ given~$Y$
indexed by unknown parameter $\psi$. (This is called a selection model
factorization; the alternative factorization into the marginal
distribution of $M$ and the conditional distribution of $Y$ given $M$ is
called a pattern-mixture model.) If $Y$ was fully observed, the
posterior distribution of the parameters would be
\[
p_{\mathrm{complete}}(\theta,\psi |Y,M) = \mathrm{const}.
\times\pi(\theta,\psi) \times L(\theta,\psi |Y),
\]
where $\pi(\theta,\psi)$ is the prior distribution of the parameters,
\[
L(\theta,\psi |Y) = f(Y|\theta) \times f(M|Y,\psi)
\]
is the complete-data likelihood and const. is a normalizing constant.
With incomplete data, the full posterior distribution becomes
\begin{eqnarray}\label{eq1}
&&p_{\mathrm{full}}(\theta,\psi |Y_{\mathrm{obs}},M)\nonumber
\\[-8pt]\\[-8pt]
&&\quad\propto\pi(\theta,\psi) \times L(\theta,\psi |Y_{\mathrm{obs}},M),\nonumber
\end{eqnarray}
where $L(\theta,\psi |Y_{\mathrm{obs}},M)$ is the observed likelihood,
obtained by integrating the missing values out of the complete-data
likelihood:
\begin{eqnarray*}
&&f(Y_{\mathrm{obs}},M|\theta,\psi)
\\
&&\quad= \int f(Y_{\mathrm{obs}},Y_{\mathrm{mis}}|\theta)f(M|Y_{\mathrm{obs}},Y_{\mathrm{mis}},\psi)\,dY_{\mathrm{mis}}.
\end{eqnarray*}
A simpler posterior distribution of $\theta$ ignores the missing data
mechanism, and is based on the likelihood given the observed
data $Y_{\mathrm{obs}}$:
\begin{eqnarray}\label{eq2}
p_{\mathrm{ign}}(\theta |Y_{\mathrm{obs}}) &\propto&\pi(\theta) \times
L(\theta |Y_{\mathrm{obs}}),
\\
L(\theta |Y_{\mathrm{obs}})
&=& \int f(Y_{\mathrm{obs}},Y_{\mathrm{mis}}|\theta)\,dY_{\mathrm{mis}},\nonumber
\end{eqnarray}
which does not involve the model for the distribution of $M$.

Statistical analysis based on (\ref{eq2}) is considerably easier than analysis
based on (\ref{eq1}), since (a) the model for the missing-data mechanism is hard
to specify, and has a strong influence on inferences; (b) the
integration over the missing data is often easier for equation~(\ref{eq2}) than
for equation~(\ref{eq1}); and (c) the full model is often under-identified or
poorly identified; identification is in some ways less of an issue in
Bayesian inference, but results rest strongly on the choice of prior
distribution. Thus, ignoring the missing-data mechanism is useful if it
is justified. Sufficient conditions for ignoring the missing-data
mechanism and basing inference on~(\ref{eq2}) (Rubin, \citeyear{Rubin1976};
Little and Rubin, \citeyear{LittleRubin2002}, Chapter~6) are as follows:
\begin{enumerate}
\item[]Missing at Random (MAR): $p(M|Y_{\mathrm{obs}},Y_{\mathrm{mis}},\psi) =
p(M|Y_{\mathrm{obs}},\psi)$ for all $Y_{\mathrm{mis}}$,

\item[]A-priori independence: $\pi(\theta,\psi) = \pi(\theta) \times\pi(\psi)$.
\end{enumerate}
Of these, MAR is the key condition in practice, and\vadjust{\eject} it implies that
missingness can depend on values in the data set that are observed, but
not on values that are missing.

The main challenges in developing posterior distributions based on (\ref{eq1})
or (\ref{eq2}) are the choice of model and computational issues, since the
likelihood based on the data with missing values is typically much more
complex than the complete-data likelihood. In the ML world, the
expectation--maximization (EM) algorithm creates a tie between the
complicated observed data likelihood and the simpler complete-data
likelihood, facilitating this computational task. Specifically, suppose
the missing-data mechanism is ignorable, and let $\theta^{(t)}$ be the
current estimate of the parameter $\theta$. The E-step of EM finds the
expected complete-data loglikelihood if $\theta$ equaled $\theta^{(t)}$:
\begin{eqnarray}\label{eq3}
&&Q \bigl( \theta |Y_{\mathrm{obs}},\theta^{(t)}  \bigr)\nonumber
\\
&&\quad= \int\log f(Y_{\mathrm{obs}},Y_{\mathrm{mis}};\theta)
\\
&&\qquad{}\cdot f \bigl(Y_{\mathrm{mis}}|Y_{\mathrm{obs}},\theta= \theta^{(t)}  \bigr) \, dY_{\mathrm{mis}}.\nonumber
\end{eqnarray}
The M-step of EM determines $\theta^{(t + 1)}$ by maximizing this
expected complete-data loglikelihood:
\begin{equation}\label{eq4}
Q \bigl( \theta^{(t + 1)}|Y_{\mathrm{obs}},\theta^{(t)}  \bigr) \ge
Q \bigl( \theta |Y_{\mathrm{obs}},\theta^{(t)}  \bigr)\quad \mbox{for all }\theta.\hspace*{-20pt}
\end{equation}

In the Bayesian world, the analog of EM is data augmentation, a variant
of the Gibbs' sampler. (An even closer analog to the Gibbs' sampler is
the Expectation Conditional Maximization algorithm, a va\-riant of EM.)
The key idea is to iterate between draws of the missing values and draws
of the parameters; draws of missing values replace expected values of
functions of the missing values in the E-step of EM, and draws of the
parameters replace maximization over the parameters in the M-step of EM.
Specifically, suppose the missing data mechanism is ignorable, and let
$( Y_{\mathrm{mis}}^{(dt)},\theta^{(\mathit{dt})})$ be current draws of the
missing data and parameters at iteration $t;$ here and elsewhere a
superscript $d$ is used to denote a draw from a distribution. Then at
iteration $t + 1$, the analog of the E-step (\ref{eq3}) of EM is to draw new
values of the missing data:
\begin{equation}\label{eq5}
Y_{\mathrm{mis}}^{(d,t + 1)} \sim p\bigl( Y_{\mathrm{mis}}
|Y_{\mathrm{obs}},\theta^{(\mathit{dt})}\bigr),
\end{equation}
and the analog of the M-step (\ref{eq4}) is to draw a new set of parameters from
the completed-data posterior distribution:
\begin{equation}\label{eq6}
\theta^{(d,t + 1)} \sim p\bigl(\theta |Y_{\mathrm{obs}},
Y_{\mathrm{mis}}^{(d,t + 1)}\bigr).
\end{equation}
As $t$ tends to infinity, this sequence converges to a draw $(Y_{\mathrm{mis}}^{(d)},\theta^{(d)})$ from the joint posterior
distribution of $Y_{\mathrm{mis}}$ and $\theta$. Those familiar with MCMC
methods will recognize this as an application of the Gibbs' sampler to
the pair of variables $Y_{\mathrm{mis}}$ and $\theta$. The utility lies
in the fact that (\ref{eq5}) is often facilitated since the distribution
conditions on the parameters, and (\ref{eq6}) is a complete-data problem since
it conditions on the imputations derived from (\ref{eq5}). For more discussion
of Bayesian computations for missing data, see Tan and Tian (\citeyear{TanTian2010}).

\section{Multiple Imputation}\label{sec:5}

Draws $Y_{\mathrm{mis}}^{(d)}$ of the missing data from equation (\ref{eq5}) at
convergence can be used to create $D > 1$ multiply-imputed data sets.
Bayesian MI combining rules can then be used for inferences that
propagate imputation uncertainty.

I outline the theory when the missing data mechanism is ignorable,
although it readily extends to the case of nonignorable nonresponse. The
idea is to relate the observed-data posterior distribution (\ref{eq1}) to the
``complete-data'' posterior distribution that would have been obtained
if we had observed the missing data $Y_{\mathrm{mis}}$, namely,
\begin{equation}\label{eq7}
p(\theta |Y_{\mathrm{obs}},Y_{\mathrm{mis}}) \propto\pi(\theta) \times
L(\theta |Y_{\mathrm{obs}},Y_{\mathrm{mis}}).
\end{equation}
Equations (\ref{eq2}) and (\ref{eq7}) can be related by standard probability theory as
\begin{eqnarray}\label{eq8}
&&p_{\mathrm{ign}}(\theta |Y_{\mathrm{obs}})\nonumber
\\[-8pt]\\[-8pt]
&&\quad= \int p(\theta|Y_{\mathrm{obs}},Y_{\mathrm{mis}})p(Y_{\mathrm{mis}}|Y_{\mathrm{obs}})\,dY_{\mathrm{mis}}.\nonumber
\end{eqnarray}
Equation (\ref{eq8}) implies that the posterior distribution $p_{\mathrm{ign}}(\theta
|Y_{\mathrm{obs}})$ can be simulated by first drawing the missing
values, $Y_{\mathrm{mis}}^{(d)}$, from their posterior distribution,
$p(Y_{\mathrm{mis}}|Y_{\mathrm{obs}})$, imputing the drawn values to
complete the data set, and then drawing $\theta$ from its
``completed-data'' posterior distribution, $p(\theta
|Y_{\mathrm{obs}},\break Y_{\mathrm{mis}}^{(d)})$. That is,
\begin{equation}\label{eq9}
p_{\mathrm{ign}}(\theta |Y_{\mathrm{obs}}) \approx\frac{1}{D}\sum_{d =
1}^{D} p \bigl(\theta |Y_{\mathrm{obs}},Y_{\mathrm{mis}}^{(d)}\bigr).
\end{equation}
When the posterior mean and variance are adequate summaries of the
posterior distribution, (\ref{eq9}) can be effectively replaced by
\begin{equation}\label{eq10}
E(\theta |Y_{\mathrm{obs}}) = E [ E(\theta
|Y_{\mathrm{obs}},Y_{\mathrm{mis}})|Y_{\mathrm{obs}}  ],
\end{equation}
and
\begin{eqnarray}\label{eq11}
\operatorname{Var}(\theta |Y_{\mathrm{obs}})
&=& E [ \operatorname{Var}(\theta|Y_{\mathrm{obs}},Y_{\mathrm{mis}})|Y_{\mathrm{obs}}  ]\nonumber
\\[-8pt]\\[-8pt]
&&{}+ \operatorname{Var} [E(\theta |Y_{\mathrm{obs}},Y_{\mathrm{mis}})|Y_{\mathrm{obs}}  ].\hspace*{-20pt}\nonumber
\end{eqnarray}
Approximating (\ref{eq10}) and (\ref{eq11}) using draws of $Y_{\mathrm{mis}}$ yields
\begin{equation}\label{eq12}
E(\theta |Y_{\mathrm{obs}}) \approx\bar{\theta} = \frac{1}{D}\sum_{d =
1}^{D} \hat{\theta} ^{(d)},
\end{equation}
where $\hat{\theta} ^{(d)} = E(\theta
|Y_{\mathrm{obs}},Y_{\mathrm{mis}}^{(d)})$ is the posterior mean of~%
$\theta$ from the $d$th completed data set, and
\begin{equation}\label{eq13}
\operatorname{Var}(\theta |Y_{\mathrm{obs}}) \approx\bar{V} + (1 + 1/D)B,
\end{equation}
say, where $\bar{V} = D^{ - 1}\sum_{d = 1}^{D} \operatorname{Var}(\theta
|Y_{\mathrm{obs}},Y_{\mathrm{mis}}^{(d)})$ is the average of the
complete-data posterior covariance matrix of $\theta$ calculated for the
$d$th data set $(Y_{\mathrm{obs}},Y_{\mathrm{mis}}^{(d)}),\break B
=\sum_{d = 1}^{D} (\hat{\theta} _{d} - \bar{\theta} )(\hat{\theta} _{d}
- \bar{\theta} )^{T} /(D - 1)$ is the between-imputation covariance
matrix, and $(1 + 1/D)$ is a~correction to improve the approximation for
small~$D$. The quantity $(1 + 1/D)B$ in (\ref{eq13}) estimates the contribution
to the variance from imputation uncertainty, missed (i.e., set to zero)
by single imputation methods.

Equations (\ref{eq12}) and (\ref{eq13}) are basic MI combining rules. Refinements that
replace the normal reference distribution for scalar $\theta$ by a
Student $t$ distribution are given in Rubin and Schenker (\citeyear{RubinSchenker1986}), with
further small-sample $t$ refinements in Barnard and Rubin (\citeyear{BarnardRubin1999});
extensions to hypothesis testing are described in Rubin (\citeyear{Rubin1987}) or
Little and Rubin (\citeyear{LittleRubin2002}, Chapter~10). Besides incorporating imputation
uncertainty, another benefit of multiple imputation is that the
averaging over data sets in (\ref{eq12}) results in more efficient point
estimates than does single random imputation. Often MI is not much more
difficult than doing a single imputation---the additional computing
from repeating an analysis $D$ times is not a major burden and methods
for combining inferences are straightforward. Most of the work is in
generating good predictive distributions for the missing values.

From a frequentist perspective, Bayesian MI for a~parametric model has
similar large-sample properties to ML, and it can be simpler
computationally. Another attractive feature of MI is that the imputation
model can differ from the analysis model by including variables not
included in final analysis. Some examples follow:
\begin{longlist}[(a)]
\item[(a)] MI was originally proposed for public use files, where the
imputer often has variables available for imputation, like geography,
that are not available\vadjust{\eject} to the analyst because of confidentiality
constraints. Such variables can be included in the imputation model, but
will not be available for analysis. In other settings, auxiliary
variables that are not suitable for inclusion in the final model, such
as side-effect data for drugs in a clinical trial, may be useful
predictors in an imputation model.

\item[(b)] For public use files, users perform analyses with different
subsets of variables. Different ML analyses involving a variable with
missing values imply different imputation models, to the extent that
they involve different sets of variables. A more coherent approach is to
multiply impute missing variables using a common model, and then apply
MI methods to each of the analyses involving subsets of variables. This
allows variables not in the subset to help predict the missing values.

\item[(c)] MI combining rules can also be applied when the complete-data
inference is not Bayesian (for example, nonparametric tests or
design-based survey inference). The assumptions contained in the
imputation model are then confined to the imputations, and with small
amounts of missing data, simple imputation models may suffice.
\end{longlist}

\section{Applications of Bayesian Methods to Missing Data Problems}\label{sec:6}

Sections~\ref{sec:4} and~\ref{sec:5} sketched the basic theory of Baye\-sian inference for
missing data and the related me\-thod of MI. We now provide some examples
of important models where these methods can be put to practical use. We
focus mainly on continuous variables, although methods for categorical
variables, and mixtures of continuous and categorical variables, are
also available (Little and Rubin, \citeyear{LittleRubin2002}).
\begin{example}[(Data with a monotone pattern of missing values)]\label{ex1}
I mentioned in Section~\ref{sec31} the factored likelihood method of Anderson
(\citeyear{Anderson1957}). Consider bivariate normal data on two variables $(Y_{1},Y_{2})$
where $Y_{1}$ is observed for all $n$ observations, and~$Y_{2}$ is
\mbox{observed} for $r < n$ observations, that is, has \mbox{$n - r$} missing values.
Assume the missing-data mechanism is ignorable. The factored likelihood
is obtained by transforming the joint normal distribution of
$(Y_{1},Y_{2})$ into the marginal distribution of $Y_{1}$, normal with
mean $\mu_{1}$ and variance $\sigma_{11}$, and the conditional
distribution of $Y_{2}$ given $Y_{1}$, normal with mean $\beta_{20 \cdot
1} + \beta_{21 \cdot 1}Y_{1}$ and variance $\sigma_{22 \cdot 1}$. The
likelihood then factorizes into the normal likelihood for $\phi_{1} =
(\mu_{1},\sigma_{11})$ based on the $n$ cases with $Y_{1}$ observed, and
the normal likelihood for $\phi_{2} = (\beta_{20 \cdot 1},\beta_{21
\cdot 1},\sigma_{22 \cdot 1})$ based on the $r$ cases with both $Y_{1}$
and $Y_{2}$ observed. The ML estimates are immediate: the sample mean
and sample variance of $Y_{1}$ (with denominator $n$) based on all $n$
observations for $\phi_{1}$, and the least squares estimates of the
regression of $Y_{2}$ on $Y_{1}$ (with no degrees of freedom correction
for $\hat{\sigma} _{22 \cdot 1}$) based on the~$r$ complete cases
for $\phi_{2}$. ML estimates of other parameters, such as the mean
$\mu_{2}$ of $Y_{2}$, are obtained by expressing them as functions of
$(\phi_{1},\phi_{2})$ and then substituting ML estimates of those
parameters. In particular, this leads to the well-known regression
estimate of $\mu_{2}$:
\begin{equation}\label{eq14}
\hat{\mu} _{2} = \hat{\beta} _{20 \cdot 1} + \hat{\beta} _{21 \cdot
1}\hat{\mu} _{1},
\end{equation}
which is easily seen to be obtained when missing values of $Y_{2}$ are
imputed as predictions from the regression of $Y_{2}$ on $Y_{1}$, with
regression coefficients estimated on the complete cases.

A corresponding Bayesian analysis is obtained by adding conjugate prior
distributions for $\phi_{1}$ and $\phi_{2}$, and computing draws from the
posterior distributions of these parameters. The posterior distribution
of $\phi_{1}$ is based on standard Bayesian methods applied to the
sample of $n$ complete observations on~$Y_{1}$---inverse-chi-squared
for~$\sigma_{11}$, normal for $\mu_{1}$ given $\sigma_{11}$, and Student's
$t$ for $\mu_{1}$. The posterior distribution of~$\phi_{2}$ is based on
standard Bayesian methods for the regression of $Y_{2}$ on $Y_{1}$
applied to the~$r$ complete observations on $Y_{1}$ and $Y_{2}$---inverse
chi-squared for~$\sigma_{22 \cdot 1}$, normal for $(\beta_{20
\cdot 1},\beta_{20 \cdot 1})$ given $\sigma_{22 \cdot 1}$, and
multivariate~$t$ for $(\beta_{20 \cdot 1},\beta_{20 \cdot 1})$. Draws
$(\mu_{1}^{(d)},\sigma_{11}^{(d)},\break\beta_{20 \cdot
1}^{(d)},\beta_{21.1}^{(d)},\sigma_{22 \cdot 1}^{(d)})$ from these
posterior distributions are simple to compute (see Little and Rubin, \citeyear{LittleRubin2002}, Chapter~7,
for details). Draws from the posterior distribution of
other parameters are then created in the same way as ML estimates, by
expressing the parameters as functions of $(\phi_{1},\phi_{2})$ and then
substituting draws. For example, a~draw from the posterior distribution
of $\mu_{2}$ is
\begin{equation}\label{eq15}
\mu_{2}^{(d)} = \beta_{20 \cdot 1}^{(d)} + \beta_{21 \cdot
1}^{(d)}\mu_{1}^{(d)},
\end{equation}
a formula that mirrors the ML formula (\ref{eq14}).

This Bayesian approach is asymptotically equivalent to ML, but it has
several useful features. First, prior knowledge about the parameters can
be incorporated in the prior distribution if this is available; if not,
noninformative reference prior distributions can be applied. Second, the
posterior distributions do a better job of capturing uncertainty in
small samples; for\vadjust{\eject} example, the draws (\ref{eq15}) incorporate $t$-like
corrections, which are not provided by standard asymptotic ML
calculations. Third, the draws yield immediate estimates of uncertainty,
such as the posterior variance, and 95\% credibility intervals. The
factored likelihood approach does not yield conveniently simple formulas
for large sample variances based on the information matrix. These are
easily approximated by draws (\ref{eq15}), and are actually superior (in a
frequentist sense) to asymptotic variances since they reflect the
uncertainty better.

Computational advantages in simulating draws\break from the posterior
distribution are modest in the current bivariate normal example, since
there are not many parameters. These benefits are more substantial in
larger problems where the factored likelihood trick can be applied.
Suppose that there are $K > 2$ variables $(Y_{1},Y_{2},\ldots,Y_{K})$ such
that (a)~the da\-ta have a~monotone pattern, such that $Y_{k}$ is always
observed when $Y_{k + 1}$ is observed, for $k = 1,\ldots,K-1;$ and (b)~%
the conditional distribution of $(Y_{k}|Y_{1},\ldots,\break Y_{k - 1})$ has a
distribution (not necessarily normal) with unknown parameters $\phi_{k}$,
for $k =1,\ldots,K;$ and (c) the parameters $(\phi_{1},\ldots,\phi_{K})$ are
distinct and have independent prior distributions. Draws from the
posterior distribution of $\phi = (\phi_{1},\ldots,\phi_{K})$ can then be
obtaining from a sequence of complete-data posterior distributions, with
the posterior distribution of~$\phi_{k}$ based on the subset of data
with $(Y_{1},\ldots,Y_{k})$ observed (Little and Rubin, \citeyear{LittleRubin2002}, Chapter~7).
This elegant scheme forms the basis for MI in the case of a monotone
pattern. In particular, SAS PROC MI yields multiple imputations for
normal models, where the regressions of $Y_{k}$ on $Y_{1},\ldots,Y_{k - 1}$
are not required to be linear and additive, as would be the case if the
joint distribution was multivariate normal.

When the data are monotone but the parameters of the sequence of
conditional distributions are not a-priori independent, or when the
pattern is not mono\-tone, these simple factored likelihood methods no
longer apply, and draws from the posterior distribution need an
iterative scheme. Markov Chain Monte Carlo methods then come to the
rescue, as in the next example.
\end{example}
\begin{example}[(The multivariate normal model with a general pattern of missing values)]\label{ex2}
Suppose observations $y_{i}$ are assumed to be
randomly sampled from a multivariate normal distribution, that is,
\[
y_{i} = (y_{i1},\ldots,y_{ip})\sim_{\mathrm{ind}}N_{p}(\mu,\Sigma),
\]
the normal distribution with mean $\mu$ and covariance matrix $\Sigma$.
There are missing values, and let~%
$y_{\mathrm{obs,}i}$, $y_{\mathrm{mis,}i}$ denote respectively the set of
observed and missing values for observation $i$. Given current draw
$\theta^{(\mathit{dt})} = (\mu^{(\mathit{dt})},\Sigma^{(\mathit{dt})})$ of the parameters, missing
values (\ref{eq5}) are drawn as
\begin{equation}\label{eq16}
y_{\mathrm{mis,}i}^{(d,t + 1)} \sim p\bigl( y_{\mathrm{mis,}i}
|y_{\mathrm{obs,}i},\theta^{(\mathit{dt})}\bigr),\quad i = 1,\ldots,n,\hspace*{-31pt}
\end{equation}
which is the multivariate normal distribution of the missing variables
given the observed variables in\break \mbox{observation} $i$, with parameters that
are functions\break of~$\theta^{(\mathit{dt})}$, readily computed using the sweep
operator (Little and Rubin, \citeyear{LittleRubin2002}, Section~7.4). New parameters (\ref{eq6}) are
drawn from the posterior distribution given the filled-in data, which is
a standard Bayesian problem, namely,
\begin{eqnarray}\label{eq17}
\Sigma^{(d,t + 1)} &\sim& p\bigl(\Sigma |Y_{\mathrm{obs}}, Y_{\mathrm{mis}}^{(d,t +
1)}\bigr),
\\\label{eq18}
\bigl(\mu^{(d,t + 1)}|\Sigma^{(d,t + 1)}\bigr)&\sim& p\bigl(\mu |\Sigma^{(d,t +1)},\nonumber
\\[-8pt]\\[-8pt]
&&\hspace*{11pt}{}Y_{\mathrm{obs}}, Y_{\mathrm{mis}}^{(d,t + 1)}\bigr),\nonumber
\end{eqnarray}
where (\ref{eq17}) is a draw from an inverse Wishart distribution, and (\ref{eq18}) is a
draw from a multivariate normal distribution. Details of these steps
were originally described in Tanner and Wong (\citeyear{TannerWong1987}) as part of their
data augmentation algorithm, and they form the basis for the multiple
imputation algorithm in SAS PROC MI,
originally developed by
Schafer\break (Schafer, \citeyear{Schafer1997}). Steps (\ref{eq16})--(\ref{eq18}) are closely related to the EM
algorithm for ML estimation, except that they lead to draws from the
posterior distribution. When feasible as here, it is recommended to
first program EM, and correct programming errors by checking that the
likelihood increases with each iteration, and then convert the EM
algorithm into the Gibbs algorithm, essentially by replacing the
conditional means of the missing values in the E-step by draws (\ref{eq16}), and
the complete-data ML parameters in the M-step by draws (\ref{eq17}) and (\ref{eq18}). MI
based on this model is available in a variety of software, including SAS
PROC MI.

A frequentist statistician might compute ML estimates and associated
standard errors based on the \mbox{information} matrix. However, the Gibbs
algorithm outlined above is simpler than computing informa\-tion-matrix
based standard errors, which are not an immediate output from EM. So a
frequentist can use the draws from Gibbs' algorithm to compute tests and
confidence intervals for the parameters, exploiting the asymptotic
equivalence of Bayes and frequentist inferences (Little and Rubin, \citeyear{LittleRubin2002},\vadjust{\eject}
Chapter~6). As in the previous example, the Bayesian approach improves
some aspects of small sample inference by including $t$-like corrections
reflecting uncertainty in the variance parameters.

Example~\ref{ex2} allows missing data to be multiply imputed for a general
pattern of missing values, rather than the monotone pattern in Example~\ref{ex1}.
A limitation is that it assumes a multivariate normal distribution
for the set of variables with missing values (normality can be relaxed
for variables that are completely observed). This is a relatively strong
parametric assumption---in particular, it assumes that the regressions
of missing variables on observed variables are normal, linear and
additive, which is not very appealing when a missing variable is binary
or regressions involve interactions, for example.

One approach to this problem is to modify the model to allow for
mixtures of continuous and categorical variables. The general location
model of Olkin and Tate (\citeyear{OlkinTate1961}) provides a useful starting point (Little
and Rubin, \citeyear{LittleRubin2002}, Chapter~14). This is useful, but the need to formulate
a tractable joint distribution for the variables is restrictive. A more
flexible approach is to \mbox{apply} MI for a sequence of conditional
regression models for each missing variable, given all the other
variables. This sequential regression multivariate imputation (SRMI)
method is only approximately Bayes, but what it loses in theoretical
coherence it gains in practical flexibility. It is the topic of the next
example.
\end{example}
\begin{example}[(Sequential regression multivariate imputation)]\label{ex3}
Suppose we have a general pattern of missing values, and
$(Y_{1},Y_{2},\ldots,Y_{K})$ are the set of variables with missing values,
and $X$ represents a set of fully observed variables. SRMI (Raghunathan
et al., \citeyear{RaghunathanEtAl2001};
Van Buuren et al., \citeyear{VanBuurenEtAl2006}) specifies models for the
distribution of each variable $Y_{j}$ given all the other variables
$Y_{1},\ldots,Y_{j - 1},Y_{j + 1},\ldots,Y_{K}$ and $X$, indexed by
parameters $\psi_{j}$, with density $g_{j}(Y_{j}|X,Y_{1},\ldots,\break Y_{j - 1},
Y_{j + 1},\ldots,Y_{K},\psi_{j})$, and a noninformative prior
distribution for $\psi_{j}$. Missing values of $Y_{j}$ at iteration $t+1$
are imputed according to the following scheme: let $y_{ji}^{(t)}$ be the
observed or imputed value of $Y_{j}$ at iteration $t$, and let
$Y^{(jt)}$ denote the filled-in data set with imputations of
$Y_{1},\ldots, Y_{j - 1}$ from iteration $t+1$ and imputations of
$Y_{j},\ldots, Y_{K}$ from iteration $t$. For $j = 1,\ldots,K$, create new
imputations of the missing values of $Y_{j}$ as follows:
\begin{longlist}[\hspace*{-6pt}]
\item[\hspace*{-6pt}] $\psi_{j}^{(t + 1)}\sim p(\psi |X,Y^{(jt)})$, the posterior distribution of
 given data  $(X,Y^{(jt)})$;

\item[\hspace*{-6pt}] $y_{ji}^{(t + 1)}\sim g_{j}(Y_{j}|X,y_{1i}^{(t + s1)},\ldots,y_{j - 1,i}^{(t + 1)},y_{j +
1,i}^{(t)},\ldots, y_{Ki}^{(t)},\break \psi_{j}^{(t + 1)})$,
 if $y_{ji}$ is missing, $i = 1,\ldots,n$.
\end{longlist}
This algorithm is repeated for $t =1,2,3,\ldots$ until the imputations
are stable; typically, more than one chain is run to facilitate
assessment of convergence (Gelman and Rubin, \citeyear{GelmanRubin1992}). The algorithm is
then repeated $D$ times to create $D$ multiply-imputed data sets, and
inferences are based on standard MI combining rules.

The positive feature of SRMI is that it reduces the multivariate missing
data problems into a set of univariate problems for each variable given
all the other variables, allowing flexibility in the choice of model for
each incomplete variable; that is, nonlinear and interaction terms are
allowed in the regressions, and the error distribution can be chosen to
match the nature of the outcome---logistic for a~binary variable, and
so on. The drawback is that the regression models for each variable
given the others does not generally correspond to a coherent model for
the joint distribution of $(Y_{1},\ldots,Y_{K})$ given $X$. Thus, the MI's
are not draws from a well-defined posterior distribution. This does not
seem to be a~major problem in practice, and SRMI is a flexible and
practical tool for handling a variety of missing data problems. Software
is available (Raghunathan, Solenberger and Van Hoewyk, \citeyear{RaghunathanSolenbergerVanHoewyk2009}; MICE, \citeyear{MICE2009}).

The regression models in SRMI are parametric, and potentially vulnerable
to model misspecification. As noted in Section~\ref{sec34}, one recent interest
in missing data research has been the development of robust methods that
do not involve strong parametric assumptions. My last example concerns
so-called ``doubly-robust methods'' for missing data.
\end{example}
\begin{example}[(Robust modeling: Penalized spline of propensity prediction)]\label{ex4}
For simplicity, we consider the case where missingness is
confined to a~single variable $Y$. Let $(x_{i1},\ldots,x_{ip},y_{i})$ be a
vector of variables for observation $i$, with $y_{i}$ observed for $i =
1,\ldots, r$ and missing for $i = r + 1,\ldots,n$, and $(x_{i1},\ldots,\break x_{ip})$
observed for $i = 1,\ldots,n$. We assume that the probability that
$y_{i}$ is missing depends on $(x_{i1},\ldots,x_{ip})$ but not $y_{i}$, so
the missing data mechanism is MAR. We consider estimation and inference
for the mean of $Y,\mu_{y} = E(Y)$. Let $m_{i}$ denote the missing-data
indicator for $y_{i}$, with $m_{i} = 1$ when $y_{i}$ is missing and
$m_{i} = 0$ when $y_{i}$ is observed.

A number of robust methods involve the propensity to be observed,
estimated by a logistic or probit regression of $M$ on $X_{1},\ldots,X_{p}$
(Rosenbaum and\vadjust{\eject} Rubin, \citeyear{RosenbaumRubin1983}; Little, \citeyear{Little1986}).
In particular, propensity
weighting computes the mean of the complete cases, weighted by the
inverse of the estimated probability that $Y$ is observed. Propensity
weighting can yield estimates with large variances, and more efficient
estimates are obtained by predicting the missing values of $Y$ based on
a model, with robustness supplied by a calibration term that weights the
residuals from the complete cases (Robins, Rotnitzky and Zhao, \citeyear{RobinsRotnitzkyZhao1994};
Rotnitzky, Robins and Scharfstein, \citeyear{RotnitzkyRobinsScharfstein1998};
Bang and Robins, \citeyear{BangRobins2005};
Scharfstein, Rotnitsky and Robins, \citeyear{ScharfsteinRotnitskyRobins1999};
Kang and Schafer, \citeyear{KangSchafer2007}). In this
context, an estimator is doubly robust (DR) if it is consistent if
either (a) the prediction model relating $Y$ to $X_{1},\ldots,X_{p}$ is
correctly specified or (b) the model for the propensity to respond is
correctly specified. In my last example, we describe a Bayesian missing
data method, called Penalized Spline of Propensity Prediction (PSPP),
that has a~DR property.

Define the logit of the propensity score for $y_{i}$ to be observed as
\begin{equation}\label{eq19}
p_{i}^{*}(\psi) = \operatorname{logit}  \bigl( \operatorname{Pr}(m_{i} =
0|x_{i1},\ldots, x_{ip};\psi)  \bigr),
\end{equation}
where $\psi$ denotes unknown parameters. The PSPP method is based on the
balancing property of the propensity score, which means the missingness
of~$y_{i}$ depends only on $(x_{i1},\ldots, x_{ip})$ only through the
pro\-pensity score, under the MAR assumption (Rosen\-baum and Rubin, \citeyear{RosenbaumRubin1983}).
Given this property, the mean of $Y$ can be written as
\begin{equation}\label{eq20}
\hspace*{20pt}\mu_{y} = E[(1 - m_{i})y_{i}] + E\bigl[m_{i} \times
E\bigl(y_{i}|p_{i}^{*}(\psi)\bigr)\bigr].
\end{equation}
Thus, the missing data can be imputed conditioning on the propensity
score. This leads to the Penalized Spline of Propensity Prediction
Method (PSPP) (Little and An, \citeyear{LittleAn2004}; Zhang and Little, \citeyear{ZhangLittle2008}). Imputations
in this method are predictions from the following model:
\begin{eqnarray}\label{eq21}
&&\bigl(y_{i}|p_{i}^{*}(\psi),x_{i1},\ldots,x_{ip};\psi,\beta,\phi,\sigma^{2}\bigr)\nonumber
\\
&&\quad\sim N\bigl(\mathrm{spl} ( p_{i}^{*}(\psi),\beta  )
\\
&&\hspace*{11pt}\qquad{}+ g(p_{i}^{*},x_{i2},\ldots x_{ip};\phi),\sigma^{2}\bigr),\nonumber
\end{eqnarray}
where $N(\nu,\sigma^{2})$ denotes the normal distribution with mean
$\nu$ and constant variance $\sigma^{2}$. The first component of the mean
function, $\mathrm{spl}( p_{i}^{*}(\psi),\beta) $, is a spline function
of the propensity score $p_{i}^{*}(\psi)$. The second component
$g(p_{i}^{*},x_{i2},\ldots, x_{ip};\phi)$ is a parametric function,
which includes any covariates other than $p_{i}^{*}$ that
predict $y_{i}$. One of the predictors, here $x_{i1}$, is omitted from the
$g$-function to avoid multicollinearity.

A variety of spline functions can be chosen; we choose a penalized
spline (Eilers and Marx, \citeyear{EilersMarx1996}; Ruppert, Wand and Carroll, \citeyear{RuppertWandCarroll2003}) of the
form
\begin{eqnarray}\label{eq22}
\mathrm{spl}(p_{i}^{*}(\psi),\beta)
&=& \beta_{0} + \beta_{1}p_{i}^{*}(\psi)\nonumber
\\[-8pt]\\[-8pt]
&&{}+ \sum_{k = 1}^{K} \beta_{k + 1} \bigl( p_{i}^{*}(\psi) - \kappa_{k}  \bigr)_{ +},\hspace*{-30pt}\nonumber
\end{eqnarray}
where 1, $p_{i}^{*}(\psi),(p_{i}^{*}(\psi) - \kappa_{1})_{ +}
,\ldots,(p_{i}^{*}(\psi) - \kappa_{k})_{ +}$ is the truncated linear basis;
$\kappa_{1} < \cdots < \kappa_{K}$ are selected fixed knots and $K$ is the
total number of knots, and $(\beta_{2},\ldots,\beta_{K + 1})$ are random
effects, assumed normal with mean 0 and variance $\tau^{2}$. This model
can be fitted by ML using a number of existing software packages, such
as PROC MIXED in SAS (SAS, \citeyear{SAS1992}; Ngo and Wand, \citeyear{NgoWand2004};
Littell et al., \citeyear{LittellEtAl1996}) and $\operatorname{lme}(\cdot)$ in S-plus (Pinheiro and Bates, \citeyear{PinheiroBates2000}). The first step
of fitting a PSPP model estimates the propensity score, for example, by
a logistic regression model or probit model of $M$ on $X_{1},\ldots,X_{p};$
in the second step, the regression of $Y$ on $P^{*}$ is fit as a spline
model with the other covariates included in the model parametrically in
the $g$-function. When $Y$ is a continuous variable we choose a normal
distribution with constant variance. For other types of data, extensions
of the PSPP can be formulated by using the generalized linear models
with different link functions.

The average of the observed and imputed values of $Y$ has a DR property,
meaning that the predicted mean of $Y$ is consistent if either (a) the
mean of $y_{i}$ given $(p_{i}^{*}(\psi),x_{i1},\ldots,x_{ip})$ in model (\ref{eq3})
is correctly specified, or (b1) the propensity $p_{i}^{*}(\psi)$ is
correctly specified, and (b2) $E(y_{i}|p_{i}^{*}(\psi),\beta) =
\mathrm{spl}(p_{i}^{*}(\psi),\beta)$.\break The robustness feature derives from
the fact that the regression function $g$ does not have to be correctly
specified, and the spline part of the regression function involves a
weak parametric assumption, practically similar to the DR property
mentioned above (Little and An, \citeyear{LittleAn2004};
Zhang and Little, \citeyear{ZhangLittle2008}).
\end{example}

How does Bayes play into these methods? The missing values of $y_{i}$
can be multiply-imputed under this model, but note that these
imputations involve a substantial number of unknown parameters, namely,
the regression coefficients and variances $(\beta,\break \tau^{2})$ of the
spline model, the regression coefficients~$\phi$ in the parametric
component $g$, the residual varian\-ce~$\sigma^{2}$, and the nuisance
parameters $\psi$ in the propensity function. Uncertainty in these
parameters is rea\-dily propagated under the Bayesian paradigm by drawing
them from their posterior distributions,\break which is reasonably
straightforward using a Gibbs' sampler.

\section{Conclusion}\label{sec:7}

I began this article by summarizing some arguments in favor of the CB
approach to statistical inference, which to my mind incorporates the
best features of both Bayesian and frequentist statistics. In short,
inferences should be Bayesian, but model development and checking
requires careful attention to frequentist properties of the resulting
Bayesian inference. This CB ``roadmap'' is not a complete solution,
since the interplay between model development and model inference,
involving questions such as what range of model uncertainty should be
included as part of formal statistical inference (Draper, \citeyear{Draper1995}), remains
illusive and hard to pin down. However, I find the CB approach a very
satisfying basis for approaching practical statistical inference.

In the remainder of the article I argued that the CB approach is
particularly attractive for dealing with problems of missing data. In a
sense, all of inferential statistics is a missing data problem, since it
involves making inferences about something that is unknown and hence
``missing''; in that broad sense, I am merely restating the previous
paragraph. However, if missing data is considered more restrictively as
referring to situations where the data matrix is incomplete, or partial
information is available on some variables, then the Bayesian paradigm
is conceptually straightforward, since likelihoods do not require a
fully-recorded rectangular data set.

Simply put, Bayesian statistics involves generating predictive
distributions of unknowns given the data. Applied to missing data, it
requires a predictive distribution for the missing data given the
observed data. Multiple imputations are simply draws from this
predictive distribution, and can be used for other analyses if a good
model is chosen for the predictive distribution.

In Sections~\ref{sec:4} and~\ref{sec:5} I outlined the basic theory underlying Bayes
inference and MI with missing data, emphasizing the role of the MAR
assumption. An important extension of this theory is to problems of
coarsened data, where some values in the data set are rounded, grouped
or censored (Heitjan and Rubin, \citeyear{HeitjanRubin1991};
Heitjan, \citeyear{Heitjan1994}). This theory has
connections with the concept of noninformative censoring that underlies
many methods of survival analysis with censored data. MI can be applied
in these settings (Hsu et al., \citeyear{HsuEtAl2006}), and is particularly useful for
conditioning imputations on auxiliary variables not included in the
primary analysis.

In Section~\ref{sec:6} I illustrated Bayesian approaches to missing data, mainly
for normal models, in view of their practical importance and historical
interest. I emphasize that Bayesian methods are also useful for
non-normal missing data problems. The SRMI methods are not restricted to
normal models, and Bayes and/or MI can be applied to categorical data
models and mixtures of categorical and continuous variables (Little and
Rubin, \citeyear{LittleRubin2002}, Chapters 13 and 14), and generalized linear models with
missing covariates (Ibrahim et al., \citeyear{IbrahimEtAl2005}). Bayesian hierarchical models
are also natural for longitudinal data (Gilks et al., \citeyear{GilksEtAl1993};
Ibrahim and Molenberghs, \citeyear{IbrahimMolenberghs2009}) and small area estimation (Ghosh and Rao, \citeyear{GhoshRao1994}), with
or without missing data.

In the examples I focused on MAR models, but Bayesian approaches to NMAR
models are also very appealing. A key problem when data are not missing
at random is lack of identifiability of the model, and Bayesian methods
provide a formal solution to the problem by allowing the formulation of
prior distributions for unidentified parameters that reflect the
uncertainty (Rubin, \citeyear{Rubin1977}; Daniels and Hogan, \citeyear{DanielsHogan2008}). Resulting inferences
are arguably superior to frequentist methods based on sensitivity
analysis, where unidentified parameters are varied over a ran\-ge. For a
recent illustration of these ideas, see Long, Little and Lin (\citeyear{LongLittleLin2010}), who
apply Bayesian missing data methods to handle noncompliance in clinical
trials.

I have focused here more on the Bayesian part of CB, given the emphasis
of the workshop that motivated this article on Bayesian tools.
Concerning the calibrated part, good frequentist properties require
realistic models for the predictive distribution of the missing values,
and this requires attention to checking the fit of the model to the
observed data, and sensitivity analyses to assess the impact of
departures from MAR. Gelman et al. (\citeyear{GelmanEtAl2005}) and Abayomi, Gelman and Levy (\citeyear{AbayomiGelmanLevy2008})
provide useful methods for model checking for multiple
imputation, and more work in this area would be useful.

\section*{Acknowledgments}

Useful comments from two referees on an earlier draft are greatly
appreciated, and helped to improve the article.

\end{document}